\documentclass[1p,numbers]{elsarticle}

\usepackage[english]{babel}
\usepackage[utf8x]{inputenc}
\usepackage{titlesec, amsmath,stackengine,graphicx,subfigure,tabularx,arydshln,url,stfloats}

\stackMath
\usepackage{multirow}
\usepackage[tableposition=top]{caption}
\usepackage[colorinlistoftodos]{todonotes}
\usepackage[left]{lineno}
\biboptions{sort&compress}
\title{Testing claims of the GW170817 binary neutron star inspiral affecting $\beta$-decay rates}
\begin{document}
\renewcommand\arraystretch{1.2}
\author[nikhef]{P.A.~Breur \corref{cor}}
\ead{sanderb@nikhef.nl}\fnref{nowat}
\author[nikhef]{J.C.P.Y.~Nobelen\corref{cor}}
\ead{jcpynobelen@gmail.com}
\author[uzh]{L.~Baudis}
\author[uzh]{A.~Brown}
\author[nikhef,uu]{A.P.~Colijn} 
\author[psi]{R.~Dressler}
\author[purdue]{R.F.~Lang}
\author[cbpf]{A.~Massafferri}
\author[cbpf]{C.~Pumar}
\author[purdue]{C.~Reuter} 
\author[psi]{D.~Schumann}
\author[freib]{M.~Schumann}
\author[asu]{S.~Towers}
\author[cbpf]{R.~Perci}

\fntext[nowat]{Now at: SLAC National Accelerator Laboratory, Menlo Park, California 94025, USA}
\cortext[cor]{Corresponding Authors}

\address[nikhef]{Nikhef and the University of Amsterdam, Science Park, 1098 XG Amsterdam, the Netherlands}
\address[uzh]{Physik-Institut, Universit\"{a}t Z\"{u}rich, 8057 Zurich, Switzerland}
\address[uu]{
Institute for Subatomic Physics, Utrecht University, Netherlands}
\address[psi]{Paul Scherrer Institut (PSI), Villingen, Switzerland}
\address[purdue]{Department of Physics and Astronomy, Purdue University, West Lafayette, IN 47907, USA}
\address[cbpf]{Centro Brasileiro de Pesquisas F\'{i}sicas - COHEP, R. Dr. Xavier Sigaud, 150 - Urca, Rio de Janeiro, Brazil}
\address[freib]{Physikalisches Institut, Universit\"{a}t Freiburg, 79104 Freiburg, Germany}
\address[asu]{Arizona State University, Tempe, AZ, USA}
\begin{abstract}
On August 17, 2017, the first gravitational wave signal from a binary neutron star inspiral (GW170817) was detected by Advanced LIGO and Advanced VIRGO. Here we present radioactive $\beta$-decay rates of three independent sources $^{44}$Ti, $^{60}$Co and $^{137}$Cs, monitored during the same period by a precision experiment designed to investigate the decay of long-lived radioactive sources. We do not find any significant correlations between decay rates in a 5\,h time interval following the GW170817 observation. This contradicts a previous claim published in this journal of an observed 2.5$\sigma$ Pearson Correlation between fluctuations in the number of observed decays from two $\beta$-decaying isotopes ($^{32}$Si and $^{36}$Cl) in the same time interval. By correcting for the choice of an arbitrary time interval, we find no evidence of a correlation above 1.5$\sigma$ confidence. In addition, we argue that such analyses on correlations in arbitrary time intervals should always correct for the so-called Look-Elsewhere effect by quoting the global significance.

\end{abstract}
	\bibliographystyle{abbrv}
\maketitle
\section{Introduction}
The first gravitational wave signal from a binary neutron star inspiral (GW170817) was detected on August 17, 2017 at 12:41:04 UTC by Advanced LIGO and Advanced VIRGO~\cite{TheLIGOScientific:2017qsa,Abbott:2018wiz}. Shortly after this observation X-ray, radio and optical counterparts were observed~\cite{Hallinan:2017woc, Troja:2017nqp, Margutti:2017cjl}. An additional claim has been made of observing a correlation of the decay rates of two $\beta$-decaying isotopes ($^{32}$Si with T$_{1/2}$ $\sim$172\,yr and $^{36}$Cl with T$_{1/2}$ $\sim$300,000\,yr), in a 5\,h time interval following the inspiral~\cite{Fischbach:2018dnd}. This correlation was attributed to a hypothesized increase in neutrino flux (created during the inspiral) at Earth. Similar claims by the same authors of a change in decay rates due to, e.g. Solar flare neutrinos~\cite{Jenkins:2008tt}, have recently been refuted~\cite{Angevaare:2018rto, Bellotti:2018jzd}. \\ \\
In this work, we present a search for correlations in the decay rates from a $\sim0.8$ kBq $^{44}$Ti\footnote{For this isotope we measure the annihilation peak of $^{44}$Sc (daughter of $^{44}$Ti), which has a half life of $\sim$4\,h.}, a $\sim0.5$ kBq $^{60}$Co and a $\sim0.8$ kBq $^{137}$Cs source, monitored by a NaI(Tl) detector setup in a 5\,h time interval following the observation of the neutron star inspiral on August 17, 2017. The detector setup~\cite{Angevaare:2018aka} located at Nikhef (Amsterdam, the Netherlands), consists of four cylindrical (76$\times$76)\,mm NaI(Tl) detector pairs shielded by 5\,cm of lead. Three detector pairs measure the de-excitation photons after the $\beta$-decay of the sources while the fourth detector pair monitors the ambient radioactive background. To obtain the number of counts contained within the full absorption peak we perform a fitting routine of three different components. Here, we consider the (Gaussian) absorption peak, the Compton spectrum, which is determined through a GEANT4 Monte Carlo (MC) simulation, and we also account for the monitored background spectrum. This routine is performed over data acquired within one-hour periods (bin size)~\cite{Angevaare:2018aka}.

\section{Methodology}
\begin{figure}[b!]
	\centering
	\captionsetup{}
	\includegraphics[width=\textwidth]{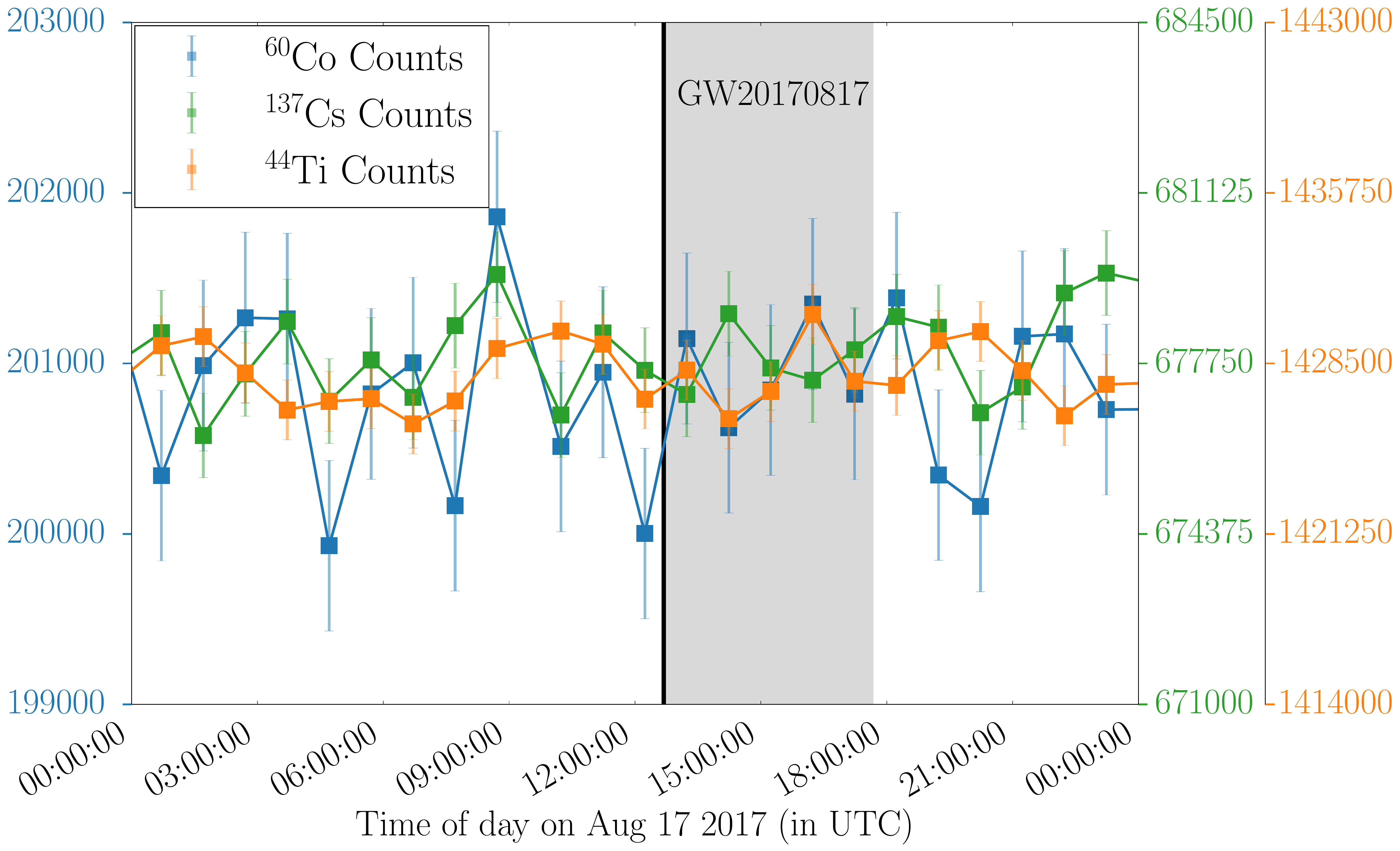}
	\caption{The number of observed $^{44}$Ti (orange), $^{60}$Co (blue) and $^{137}$Cs (green) decays per hour as recorded on August 17, 2017. The grey-shaded region denotes the 5\,h following the neutron star inspiral observation (black line). The depicted errors are of statistical origin and the ranges of the y-axis are set to $\pm1$\% of the mean decay rate for each source. }
	\label{fig:sources_during_inspiral}
\end{figure}

Figure~\ref{fig:sources_during_inspiral} shows the monitored radioactive decay rates of $^{44}$Ti (orange), $^{60}$Co (green) and $^{137}$Cs (blue) as a function of time on August 17, 2017. We also indicate the 5\,h time interval (grey area) following the neutron star inspiral observation (black line). Due to this short measuring period compared to the half-life of the isotopes, no correction is made for the expected exponential decrease in activity. Because there is no known physics model proposing a correlation between gravitational waves and changes in radioactive decay rates, we follow the analysis method of~\cite{Fischbach:2018dnd}. If we would assume such an underlying physical mechanism to exist, describing the influence of neutrinos on the decay rates of $\beta$ emitters, we should find any effect equally in magnitude for all sources decaying through the $\beta$ mechanism, as argued in~\cite{Jenkins:2012pb}. \\ \\
To test for a hypothetical correlation between the decay rates of $^{44}$Ti, $^{60}$Co and $^{137}$Cs during the observation of the neutron star inspiral signal, we use the Pearson Correlation (PC) test~\cite{Pearson}. The PC test returns a test statistic \textit{r}, which indicates the measure of correlation between the measured decay rate of two sources ($x$ and $y$) and is defined as
\begin{equation}
r_{xy} = \frac{1}{N}\frac{\sum_{i=1}^{N}(x_i-\bar{x})(y_i-\bar{y})}{\sigma_x \sigma_y},
\label{eqn:pctest}
\end{equation}
where $x_i(y_i)$ are the number of observed decays per bin $i$. $\bar{x}(\bar{y})$ and $\sigma_x(\sigma_y)$ are the expectation values and standard deviations, respectively, for the $N$-sized sample of the two radioactive sources. The test statistic \textit{r} is by definition a number between $-1$ (negative correlation) and 1 (positive correlation). The local test statistic $r_{xy}$ is used in our two-sided test to find a significance ($p$-value) of the correlation, which we define as $p_{xy}$.\\
Since a physical mechanism that causes a correlation as claimed by~\cite{Fischbach:2018dnd} is unknown, there are different time intervals in which one may test for a correlation. Because the selection of time intervals can influence the significance of a correlation, the Look-Elsewhere Effect~\cite{Lyons:1900zz} must be taken into account by referencing the global significance.\\ \\
In this work, we apply a similar methodology to determine such a global significance as in~\cite{Angevaare:2018rto}. We define five time intervals as $[t_0,t_0 + t_i]$, where $t_0$ is the time of the GW170817 observation and $t_i \in [3,8]$ in hours. A MC simulation of the experiment is conducted with 30,000 trials. 
The sample size is chosen such that the expected statistical power approximates 1 for the reported effect size of r = 0.950 and up to confidence levels of 3$\sigma$ ($\alpha = 0.003$). Data points are drawn from a Poisson distribution under the null hypothesis (exponential decay without any systematic influence). For every trial, a PC test is performed on every considered time interval and on each combination of the three sources. Subsequently, we define a new test statistic, based on the p-value of most extreme correlation per trial:
\begin{equation}
\setstackgap{S}{2pt}
\overline{p}_{xy} = \stackunder{\stackon{\text{min}}{\scriptstyle t_f}}{\scriptstyle t_i}(p_{xy}),
\label{eqn:pc_correction}
\end{equation}
where $t_i$ and $t_f$ are the boundaries of the chosen time interval and $\overline{p}_{xy}$ is the statistic of the local PC test on two radioactive decay rates of sources \textit{x} and \textit{y}. From the distribution of this test statistic under the null hypothesis, we determine the p-value of observing at least one such extreme correlation during the trial when the null hypothesis is true. \\
We perform the PC test over three different combinations of sources. For every trial in the MC, we thus also return the statistic of the most significant correlation regardless of source combination and the chosen time interval (between $t_i$ and $t_f$),
\begin{equation}
\setstackgap{S}{2pt}
\widetilde{p} = \stackunder{\stackon{\text{min}}{\scriptstyle t_f}}{\scriptstyle t_i}(p_{xy},p_{xz},p_{yz}),
\label{eqn:teststatistic_mc}
\end{equation}
where $p_{xy}$, $p_{xz}$ and $p_{yz}$ are the local PC test (Eq.~(\ref{eqn:pctest})) $p$-values calculated for the five defined time intervals per arbitrary combination of sources \textit{x}, \textit{y} and \textit{z}. \\
Finally, by computing the distribution of extreme statistics found in Eq.~(\ref{eqn:pc_correction}) and~(\ref{eqn:teststatistic_mc}), we assess the $p$-values and retrieve two unique global significances, which are used to correct the local correlation significance: once for the arbitrary choice of the 5\,h time interval, and once for the choice of the 5\,h time interval while monitoring three sources, respectively.

\section{Results}
Table \ref{tab:Allcorr} shows the results for the three different combination of sources. We find uncorrected (local) correlations in the 5\,h interval following the GW170817 observation at 2.4$\sigma$ ($p = 0.015$), 1.3$\sigma$ ($p = 0.19$) and 1.8$\sigma$ ($p = 0.07$) significance for $^{44}$Ti$-^{60}$Co, $^{44}$Ti$-^{137}$Cs and $^{60}$Co$-^{137}$Cs, respectively. It is important to stress that these local significances cannot be interpreted as a valid result to reject the null hypothesis, and can be straightforwardly explained by expected stochastic variation in the data.\\
\begin{table*}[tb!]
	\centering
	\begin{tabular}{c  ||c|  c  c}
		
		\small\text{Isotope}& \text{Local $\sigma$} & \multicolumn{2}{c}{\text{Global } $\sigma$}  \\
		\small\text{combination}&(Uncorrected)&\multicolumn{2}{c}{(Corrected)}  \\ \hline
		& &\small\text{Period} &\small\text{Period +} \\
		&&& \small{\text{Source}} \\  \hline \hline
		
		{$^{44}$Ti - $^{60}$Co} 
			& 2.4  					
			&  1.9 								
			&\text{1.4} \\					

		{$^{44}$Ti - $^{137}$Cs }
		& 1.3										
		& 0.8               					 	
		& 0.2 \\									
		
		$^{60}$Co - $^{137}$Cs 
		&1.8 										
		&1.3										
		&0.7 \\										

	\end{tabular}
	\caption{Results of the Pearson Correlation (PC) test on the measured radioactive decay rates for all combinations of sources ($^{44}$Ti, $^{60}$Co and $^{137}$Cs) in the 5\,h interval following the GW170817 observation. We correct for the arbitrary choice of the time interval (`Period') with the global significance computed from Eq.~(\ref{eqn:pc_correction}). Finally, we also correct for monitoring three sources at the same time (`Period + Source') with the global significance computed from Eq.~(\ref{eqn:teststatistic_mc}).}\label{tab:Allcorr}
\end{table*}\\ After correcting for the choice of time intervals, Eq.~(\ref{eqn:pc_correction}), and the monitoring of multiple sources, Eq.~(\ref{eqn:teststatistic_mc}), we find correlations of 0.2$\sigma$, 0.7$\sigma$ and 1.4$\sigma$ for the three isotope combinations. The most extreme correlation at 1.4$\sigma$ ($p = 0.15$) significance is between the decay rates of $^{44}$Ti and $^{60}$Co. This result shows that the neutron star inspiral does not lead to a significant correlation between the decay rates of independent $\beta$-decaying isotopes and that the null hypothesis can thus not be rejected. \\ \\
Experimental conditions such as the temperature, photomultiplier tube high voltage, magnetic field strength and radon activity were continuously monitored during the measurement period presented in Figure~\ref{fig:sources_during_inspiral}. Variations in these parameters are within operating range as defined by Table 4 in~\cite{Angevaare:2018aka}. The pressure and humidity were also continuously monitored and remained stable during the full measurement period at the 1\% level with a 68\% confidence level. We conclude that the systematic influences due to these experimental conditions remain below $ 2\times10^{-5}$. This is more than an order of magnitude smaller than the observed absolute rate differences in counts between individual data points in Figure~\ref{fig:sources_during_inspiral}, showing that a possible effect has not been significantly enhanced (or diminished) by changes in the experiment's environment.\\

\section{Discussion}
Although the results of this work show no statistically significant evidence for a correlation in the radioactive decay rate post inspiral, the claim of such a correlation by~\cite{Fischbach:2018dnd} needs to be discussed.\\ \\
First, the reported correlation at 2.5$\sigma$ confidence in~\cite{Fischbach:2018dnd} was not corrected for the arbitrary choice of a 5\,h time interval. We conclude that the claimed effect is consistent with expected stochastic variation in the data and a further analysis, including a MC simulation\footnote[1]{From the published data~\cite{Fischbach:2018dnd} we find that the claimed 2.5$\sigma$ decreases to 1.9$\sigma$ when corrected for the choice of time interval using a MC.} should lead to a similar conclusion. The choice of an arbitrary, 5-hour period after the merger observation can greatly influence the significance of a correlation as is shown in our work. Furthermore, the arbitrary choice of one-hour bin-sizes and its possible influences on the global significance were not calculated and accounted for. It was shown in~\cite{Towers:2012jq} and \cite{towers:2018} that the process of fitting with variable binning can produce biased estimates, which may result to the incorrect acceptance of a hypothesis.\\ \\ 
The claim of a correlation in~\cite{Fischbach:2018dnd} was thus not properly corrected for the arbitrary choice of a time interval. This led to a problematic interpretation of the data. In the future, we urge the authors of~\cite{Fischbach:2018dnd} to always define a model to test prior to un-blinding the data to account for any human bias. We urge the reader to be critical towards presented results without any correction for such biases.

\section{Conclusion}
We have found no statistically significant evidence of a correlation between the radioactive decay rate of $\beta$-decaying isotopes $^{44}$Ti, $^{60}$Co and $^{137}$Cs in a 5\,h time interval following the observation of the neutron star inspiral (GW170817). Correlations corrected for the choice of time intervals are within 1.5$\sigma$ confidence for all source pairs. The largest observed correlation was found for the decay rates of $^{44}$Ti and $^{60}$Co at 1.4$\sigma$ significance, which also includes a correction for finding spurious correlations while monitoring three sources simultaneously. Thus, we cannot reject the null hypothesis that describes exponential $\beta$-decay.

\section*{Acknowledgements}
We are grateful for the support of the Purdue Research Foundation, the Netherlands Organization for Scientific Research (NWO), the University of Zurich, the Swiss National Science Foundation under Grant Nos. 200020-162501 and the Foundation for Research Support of the State of Rio de Janeiro (FAPERJ).

\end{document}